\documentclass[twocolumn,prb,showpacs,superscriptaddress]{revtex4-1}

\usepackage{epsfig}
\usepackage{graphics}% Include figure files
\usepackage{graphicx}% Include figure files
\usepackage{dcolumn}% Align table columns on decimal point
\usepackage{bm}% bold math
\usepackage{amssymb}
\usepackage{amsmath}
\usepackage{verbatim}
\usepackage{color}

\begin{document}

\title
{
Einstein Modes in the Phonon Density of States of the Single-Filled Skutterudite Yb$_{0.2}$Co$_{4}$Sb$_{12}$ 
}

\author{I. K. Dimitrov}
\email[Corresponding author.  Electronic address:]{\newline idimitrov@bnl.gov}
\affiliation{Condensed Matter Physics and Materials Science Department, Brookhaven National Laboratory, Upton, New York 11973-5000, USA}

\author{M. E. Manley}
\affiliation{Lawrence Livermore National Laboratory, Livermore, California 94551, USA}

\author{S. M. Shapiro}
\affiliation{Condensed Matter Physics and Materials Science Department, Brookhaven National Laboratory, Upton, New York 11973-5000, USA}

\author{J. Yang}
\affiliation{State Key Laboratory of High Performance Ceramics and Superfine Microstructure, Shanghai Institute of Ceramics, Chinese Academy of Sciences, Shanghai 200050, China}

\author{W. Zhang}
\affiliation{State Key Laboratory of High Performance Ceramics and Superfine Microstructure, Shanghai Institute of Ceramics, Chinese Academy of Sciences, Shanghai 200050, China}

\author{L. D. Chen}
\affiliation{State Key Laboratory of High Performance Ceramics and Superfine Microstructure, Shanghai Institute of Ceramics, Chinese Academy of Sciences, Shanghai 200050, China}

\author{Q. Jie}
\affiliation{Condensed Matter Physics and Materials Science Department, Brookhaven National Laboratory, Upton, New York 11973-5000, USA}

\author{G. Ehlers}
\affiliation{Spallation Neutron Source, Oak Ridge National Laboratory, Oak Ridge, Tennessee 37831-6475, USA}

\author{A. Podlesnyak}
\affiliation{Spallation Neutron Source, Oak Ridge National Laboratory, Oak Ridge, Tennessee 37831-6475, USA}

\author{J. Camacho}
\affiliation{Condensed Matter Physics and Materials Science Department, Brookhaven National Laboratory, Upton, New York 11973-5000, USA}

\author{Qiang Li}
\email[Corresponding author.  Electronic address:]{\newline qiangli@bnl.gov}
\affiliation{Condensed Matter Physics and Materials Science Department, Brookhaven National Laboratory, Upton, New York 11973-5000, USA}

\date{\today}


%%%%%%%%%%%%%%%%%%%%%%%%%%%%%%%%%%%%%%%%%%%%%%%%%%%%%%%%%%%%%%%%%%%%%%%%%%%%%%%%%%%%%%%%%
\begin{abstract}

Measurements of the phonon density of states by inelastic neutron \emph{time-of-flight} scattering and specific heat measurements along with first principles calculations, provide compelling evidence for the existence of an Einstein oscillator ("\emph{rattler}") at ${\omega}_{E1}  \approx$ 5.0 meV in the filled skutterudite Yb$_{0.2}$Co$_{4}$Sb$_{12}$.  Multiple dispersionless modes in the measured density of states of Yb$_{0.2}$Co$_{4}$Sb$_{12}$ at intermediate transfer energies (14 meV $\leq$ \emph{$\omega$} $\leq$ 20 meV) are exhibited in both the experimental and theoretical \emph{density-of-states} of the Yb-filled specimen.  A peak at 12.4 meV is shown to coincide with a second Einstein mode at \emph{$\omega_{E2} \approx$} 12.8 meV obtained from heat capacity data.  The local modes at intermediate transfer energies are attributed to altered properties of the host CoSb$_{3}$ cage as a result of Yb-filling.  It is suggested that these modes are owed to a complementary mechanism for the scattering of heat-carrying phonons in addition to the mode observed at ${\omega}_{E1} \, \approx$ 5.0 meV.  Our observations offer a plausible explanation for the significantly-higher \textit{dimensionless figures of merit} of filled skutterudites, compared to their parent compounds.

\end{abstract}

%%%%%%%%%%%%%%%%%%%%%%%%%%%%%%%%%%%%%
\maketitle

\section{Introduction}

Skutterudites, RM$_{4}$Sb$_{12}$ (R = rare earth element; M = Fe or Co), are an important class of materials for studies of fundamental physics problems, since some of them exhibit unconventional superconductivity,\cite{Kawasaki} while others show heavy fermion behaviors.\cite{Yuhasz}  In the applied area, certain types of filled skutterudites, such as iron and cobalt antimonides, hold the promise of becoming excellent candidates for real-life thermoelectric applications, particularly in the 400 $-$ 800 K range.  In the mid-90's Slack proposed the \emph{phonon glass-electron crystal} (PGEC) model,\cite{Slack} which established the lattice thermal conductivity, ${\kappa}_{L}$, as an important tuning parameter in the quest for thermoelectric materials with high \textit{dimensionless figures of merit}, \emph{ZT}.\cite{Sales}  The filled skutterudites have been shown to have much lower ${\kappa}_{L}$'s than their parent compounds, since heavy atoms, such as La, Ce, Yb or Tl fill the cages of the open host structure and are expected to exhibit localized atomic vibrations, called "rattlers", which scatter heat-carrying acoustic phonons.  The presence of such localized oscillations superimposed on collective motions of plane wave modes are among a few known examples of Einstein modes in nature.  The "rattler mode" scenario seemed to emerge as the correct model for explaining experimental neutron time-of-flight scattering data and specific heat measurements in the filled skutterudites,\cite{Mandrus,Keppens,Bauer,Hermann} thus providing a plausible explanation for the much lower lattice thermal conductivities in the latter compared to their parent compounds.

However, the findings of a recent inelastic neutron experiment on Ce and La-filled skutterudites suggested that the guest eigenmodes are coupled to lattice degrees of freedom.\cite{Koza}  At the same time, some very recent theoretical calculations by Veithen and Ghosez\cite{Veithen} have suggested a possible coupling of guest and host dynamics at low transfer energies in the densities of states in La-filled skutterudites, consistent with the findings of Koza \textit{et al.}.\cite{Koza}  Furthermore, the same authors have suggested that heavier elements, such as Tl, may possess single Einstein modes in the density of states, while arguing that what has been interpreted as a second guest oscillator (\emph{i.e.} ${\Theta}_{E2} \approx$150$-$200 K),\cite{Mandrus,Keppens} could be due to coupled lattice vibrations.\cite{Veithen}  Therefore, understanding the physical basis of the second Einstein mode, \emph{${\Theta}_{E2}$}, is relevant for understanding the large differences in the \emph{ZT}'s between filled skutterudites and their parent compounds.

In this article, we aim to contribute to the understanding of filled skutterudite thermoelectrics by comparing the experimentally-measured density of states (DOS) of Yb$_{0.2}$Co$_{4}$Sb$_{12}$ from time-of-flight inelastic neutron scattering data with a theoretical Density Functional Perturbation Theory (DFPT) calculation of the DOS in Yb$_{0.25}$Co$_{4}$Sb$_{12}$, and correlating the Einstein temperatures obtained from a heat capacity measurement on Yb$_{0.2}$Co$_{4}$Sb$_{12}$ with local modes observed in both the theoretical and experimental DOS's.  

The paper is organized as follows.  In Sec. II we discuss the materials preparation methodology, review the neutron time-of-flight instrumentation and data analysis utilized in the derivation of the neutron-weighted density of states of Yb$_{0.2}$Co$_{4}$Sb$_{12}$, the theoretical approach for the derivation of the density-functional calculation of the DOS in Yb$_{0.25}$Co$_{4}$Sb$_{12}$, and finally, the measurement technique of the heat capacity of Yb$_{0.2}$Co$_{4}$Sb$_{12}$ with an outline of the procedure for obtaining the Einstein temperatures of the studied sample.  The main findings of the joint experimental and theoretical work are given in Sec. III.  Finally, we conclude with a summary in Sec. IV. 

%%%%%%%%%%%%%%%%%%%%%%%%%%%%%%%%%%%%%%%%%%%%%%%%%%%%%%%%%%%%%%%%%%%%%%%%%%%%%%%%%%%%%%%%%%%%%%%%%%%

\section{Materials and Methods}

\subsection{Materials Synthesis}

A polycrystalline Yb$_{0.2}$Co$_{4}$Sb$_{12}$ sample was prepared \textit{via} a solid state method.  High-purity Yb (pieces 99.9$\%$, Alfa Aesar), Co (Cobalt Slut 99.95 $\%$, Alfa Aesar) and Sb (shots, 99.9999$\%$, Alfa Aesar) were mixed in stoichiometric quantities.  The mixture was sealed in a carbon-coated quartz tube under 300 mbar Ar pressure.  The quartz ampoule was heated to 1000$^{\circ}$C at 2.5$^{\circ}$C/min, then slowly heated to 1100$^{\circ}$ (at $\approx$ 1$^{\circ}$C/min) and left for about 30 h.  Subsequently, the ampoule with the homogeneous molten liquid was removed from the furnace at 1100$^{\circ}$C and quenched into a water bath.  The ingot which formed as a result of the quench was placed in a furnace and annealed at 680$^{\circ}$C for 168 h in order to form the correct crystallographic phase.  Subsequently, it was powdered and spark-plasma sintered into bulk pellets at 50 MPa and 620$^{\circ}$C for 2 minutes in a protective Ar atmosphere.  The high density ($>$ 99.5$\%$ theoretical density) pellets consist of \emph{single phase} Yb$_{0.2}$Co$_{4}$Sb$_{12}$ compound, as confirmed by X-ray diffraction and transmission electron microscopy analysis.

\subsection{Inelastic Neutron Time-of-Flight Measurements and Analysis}
  
The inelastic neutron \emph{time-of-flight} neutron scattering studies were performed on the Cold Neutron Chopper Spectrometer (CNCS) of the Spallation Neutron Source (SNS) at Oak Ridge National Laboratory.  The sample ($\approx9.15$ g of polycystalline millimeter-sized pieces) was loaded into an aluminum housing fastened to a suspended aluminum frame.  A cadmium mask was placed in front of the holder exposing only the sample to the incident beam in order to reduce the background signal.  Prior to the scattering event, the neutrons were selected by a high-speed double disk chopper with sensitive phase adjustment operated in the high-resolution mode of the instrument.  Subsequently, the scattered neutrons were collected by detectors placed 3.5 m from the sample.  Neutrons with incident energies of 9.09 meV provided for a usable range from --5 to approximately 45 meV on the neutron energy gain side.  The energy resolution at zero energy transfer was 0.186 meV at FWHM.  The sample and the empty aluminum can were each measured at ambient temperature ($\approx$ 300 K) for 8 h in order to obtain a satisfactory signal-to-noise ratio in the measured intensities.  

The experimental results were analyzed as follows: The recorded neutron counts were normalized with respect to proton charge on target and sorted in 1 $\mu$sec bins.  Subsequently, the dynamic structure function $S(q,\omega)$ was integrated in the $q$-range from $3-6$ ${\buildrel_{\circ} \over {\mathrm{A}}}^{-1}$.  The background spectrum, comprising of scattering through an empty aluminum holder, ${{S}^{b}(\omega)}$, was subtracted from the sample spectrum, $S(\omega)$, and converted into a precursor to the density of states as follows:

\begin{equation}
 	G^{p}(\omega) = \frac{\left[ S\left( \omega\right) - {S}^{b}\left( \omega\right) \right] \cdot \omega }{ \coth\left( \frac{\omega \cdot \beta}{2}\right) - 1}
\end{equation}
, where $G^{p}(\omega)$ and $\omega$ stand for precursor to the density of states (in arbitrary units) and the neutron energy gain (in meV), respectively, while $\beta = {\left ({k}_{B} T \right)}^{-1}$.\cite{Squires}  ${G}^{p}(\omega)$ is the neutron-weighted density of states and it deviates from the true density of states ${G}(\omega)$ for a number of reasons.  Most notably, multiphonon scattering contributions are also embedded in ${G}^{p}(\omega)$, which need to be removed in order to obtain the true density of states $G(\omega)$.\cite{Ashcroft,Squires,Fultz}

\begin{figure}[t]
 	\begin{center}
		\includegraphics[width=0.45\textwidth]{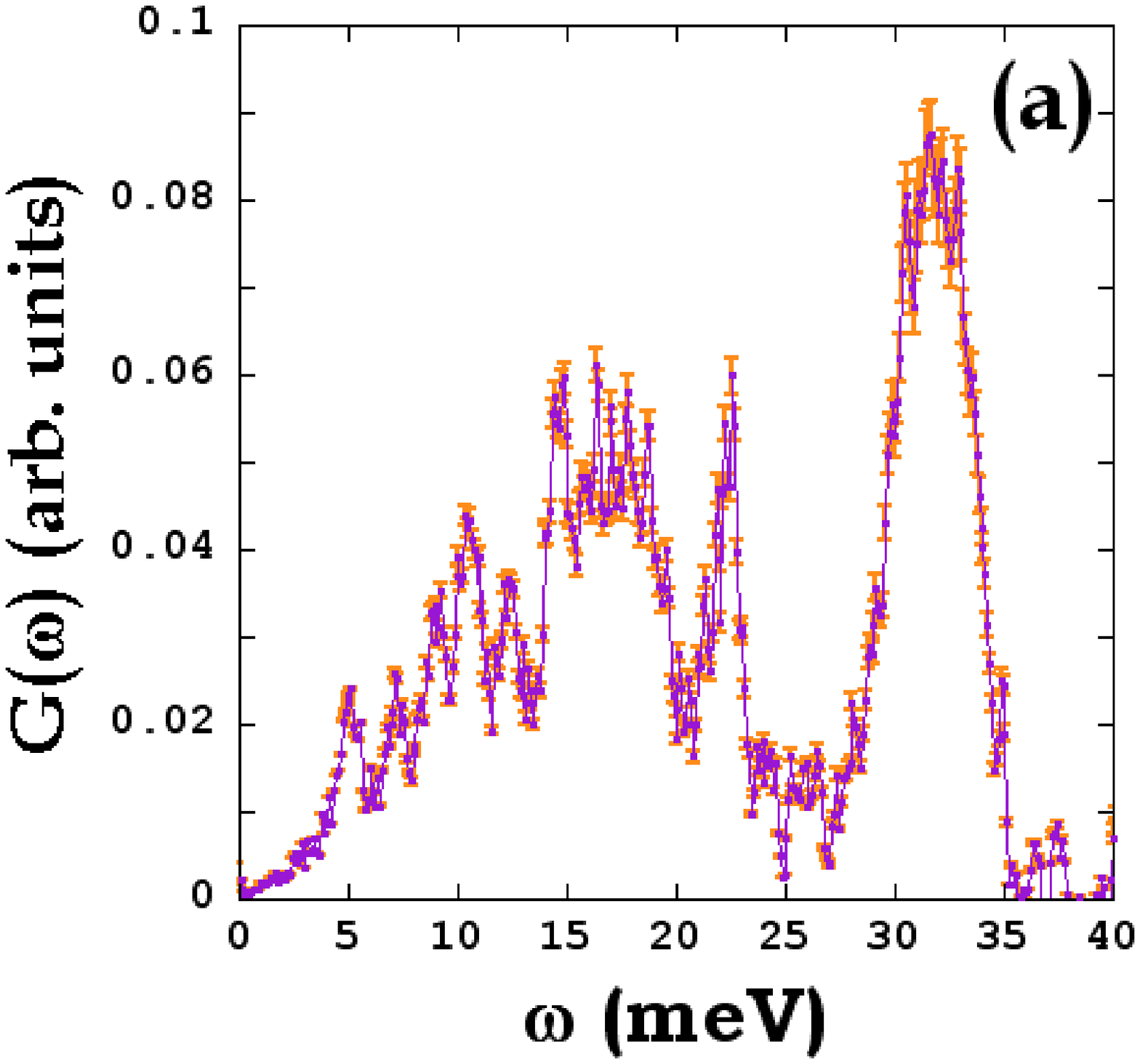}\vspace{0.5cm}
		\hspace{-1.2cm}
		\vspace{0.2cm}
		\includegraphics[width=0.45\textwidth]{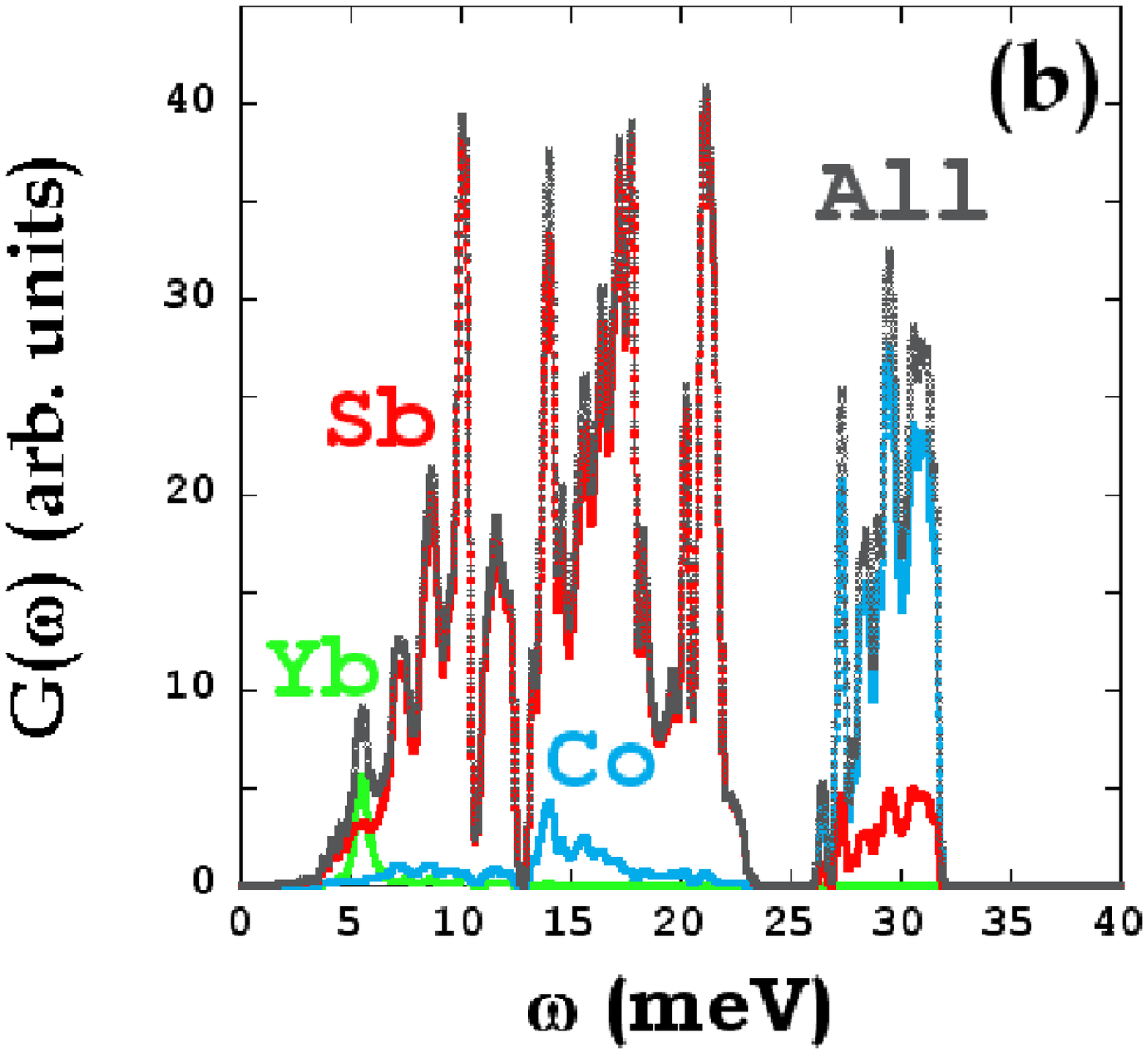}
		\caption{(color online).  Shown is the remarkable correspondence between (a) the \emph{G($\omega$)} of Yb$_{0.2}$Co$_{4}$Sb$_{12}$ derived from the inelastic neutron time-of-flight neutron scattering experiment and (b) the results of the density functional perturbation calculation of the phonon DOS in Yb$_{0.25}$Co$_{4}$Sb$_{12}$ (plotted in gray).  Individual contributions of the different elements are as shown: Sb - red, Yb - green and Co - blue (b).  The experimental data in (a) exhibit modes which account for those predicted by the theory.  A clearly-defined mode is visible at (a) $\omega \approx$ 5 meV, agreeing with (b) a distinct peak at $\omega$ $\approx$ 5.52 meV in the theoretical $G(\omega)$.}
	\end{center}
\end{figure}

A number of techniques for removing multiphonon scattering from $G^{p}(\omega)$ have been employed.\cite{Fultz}  These range from the simplest removal of a constant, to treating the multiphonon part in the incoherent approximation.\cite{Fultz,Sears,Kresch}  Here, we calculated the multiphonon scattering to all orders in the incoherent approximation using a phonon expansion procedure described in detail elsewhere.\cite{Manley}  Not knowing the phonon DOS \textit{a priori}, the phonon expansion was calculated using a theoretical partial phonon DOS calculated for each atom type from first principles.  The multiphonon scattering contributes about 2$\%$ relative to the 1-phonon scattering and was mostly flat.  Since the multiphonon contribution was small and since the final phonon DOS was fairly similar to the theoretical phonon DOS, no further iterations were needed.  In order to remove the residual background signal from air scattering, a third order polynomial was fitted through the low-energy (0 $-$ 4 meV) and high energy (37.55 $-$ 44.95 meV) tails of $G^{p}(\omega)$ and subsequently subtracted from ${G}^{p}(\omega)$ in order to obtain $G(\omega)$.  The density of states is expected to vary as ${\omega}^{2}$ at the lowest transfer energies 0 $\leq$ $\omega$ $\leq$ 3 meV, and vanish above the high-energy optic modes ($\omega$ $\geq$ 36 meV), providing the boundary conditions for the fit.  Finally, $G(\omega)$ was normalized and the results displayed in Fig. 1(a).  

The experimental density of states (see Fig. 1(a)) was used for the derivation of the sample specific heat by performing the discrete sum over the energy exchange, ${\omega}_{i}$, according to:

\begin{equation}
	{C} = \left( \frac{{k}_{B}}{\hbar} \right) \frac{1}{{\left( {k}_{B} T \right)}^{2}} \sum_{i}  \frac{{ {\omega}_{i} }^{2} \, {e}^{{\omega}_{i}/{k}_{B} T}}{{\left ( {e}^{{\omega}_{i}/{k}_{B} T} - 1 \right)}^{2}} \, G \left( {\omega}_{i} \right) \,  \Delta \omega  
\end{equation}
, where $\hbar$ and $\Delta \omega$ stand for Planck's constant and the discrete energy separation between data points ($\Delta \omega$ = 0.1 meV), respectively.  The heat capacity derived from the DOS ($G \left( \omega \right)$) offers a plausible consistency check on the results of the direct measurement.  The results of the conversion are exhibited in Fig. 2. 

%%%%%%%%%%%%%%%%%%%%%%%%%%%%%%%%%%%%%%%%%%%%%%%%%%%%%%%%%%%%%%%%%%%%%%%%%%%%%%%%%%%%%%%%%%%%%%%%%%%%%%%%%

\subsection{Theoretical Approach}

The "Frozen Phonon" method\cite{Martin} was adopted in order to obtain the lattice dynamic properties of Yb-filled skutterudites.  We started from a totally-optimized cell of Yb$_{0.25}$Co$_{4}$Sb$_{12}$ (2 $\times$ 2 $\times$ 2 supercell of Co$_{4}$Sb$_{12}$ together with 2 Yb atoms) by using rigorous computational parameters: 1$\times$10$^{-8}$ eV for self-consistent calculation and 1$\times$10$^{-4}$ eV/${\buildrel_{\circ} \over {\mathrm{A}}}$ for Hellmann-Feynman force convergence.  Then, the atoms were given small finite displacements ${U}_{0} = \pm 0.01 \, {\buildrel_{\circ} \over {\mathrm{A}}}$, and the number of displacements was determined by the symmetry of the system.  All the energy and force calculations were carried out under the projector augmented wave (PAW) method.\cite{Blochl,Kresse1} implemented in the Vienna \textit{ab initio} simulation package (VASP).\cite{Kresse2}

The force constant matrix elements are the second derivatives of the total energies with respect to the pairwise finite displacements.  The dynamic matrix \textbf{D(q)}, where \textbf{q} is a phonon wave vector in the first Brillouin zone, was derived by taking the Fourier transform of the force constant matrix.  The square of the phonon frequency $\omega$(\textbf{q},$\nu$) and the eigenvector \textbf{e}(\textbf{q},$\nu$) were computed by solving the following equation:

\begin{equation}
\textbf{D} (\textbf{q} ) \cdot \textbf{e} (\textbf{q},\nu) = {\omega}^{2} ( \textbf{q},\nu ) \; \textbf{e}  (\textbf{q},\nu )  
\end{equation}
, where $\nu$ is the index of vibrational modes.  Then, the partial phonon density of states was calculated by:

\begin{equation}
{G}_{i,\mu}(\omega) = \frac{1}{n} \sum_{\textbf{q},\nu} {|{e}_{i,\mu}(\textbf{q},\nu)|}^{2} g \left( \omega - \omega(\textbf{q},\nu) \right)
\end{equation}
, where \emph{n}, \emph{i} and \emph{$\mu$} indicate the number of \emph{q} points, atom number and fraction of three-directional parts, respectively.  In our calculation for Yb$_{0.25}$Co$_{4}$Sb$_{12}$, we used a Gauss broadening function \emph{g($\omega$)} with a smearing factor of 0.1.  The resulting DOS is shown in Fig. 1(b).

%%%%%%%%%%%%%%%%%%%%%%%%%%%%%%%%%%%%%%%%%%%%%%%%%%%%%%%%%%%%%%%%%%%%%%%%%%%%%%%%%%%%%%%%%%%%%%%%%%%%%%%%%

\subsection{\emph{in-situ} Heat Capacity Studies of Yb$_{0.2}$Co$_{4}$Sb$_{12}$}

The heat capacity, ${C}_{p}$, of Yb$_{0.2}$Co$_{4}$Sb$_{12}$ was measured on a Physical Properties Measurement System unit by Quantum Design, utilizing its Heat Capacity Option.  A small piece of material was removed from the same experimental sample for the inelastic neutron time-of-flight measurement, which was subsequently cut and polished to a parallelepiped (\emph{m} = 0.0244 g; 2.195 mm$\times$1.917 mm$\times$0.931 mm).  

In order to assess the relative contributions of lattice phonons and "rattler" Yb atoms, we measured the heat capacity of the Yb-filled sample with respect to temperature.  The focus was predominantly on the low temperature range of the data (6 K $\leq$ $T$ $\leq$ 53.5 K), such that the experiment would probe low-energy heat-carrying phonons without the obscuring influences of high-energy phonons and anharmonic effects.\cite{Ashcroft}  The heat capacity, \emph{C$_{p}$}, of Yb$_{0.2}$Co$_{4}$Sb$_{12}$ was divided by the temperature, \emph{T}, and plotted as a function of ${T}^{2}$.  The results are displayed in Figure 2 in J/K$^{2}$-mole of atoms (plotted with orange circular signs).  

\begin{figure}
 	\includegraphics[width=0.46\textwidth]{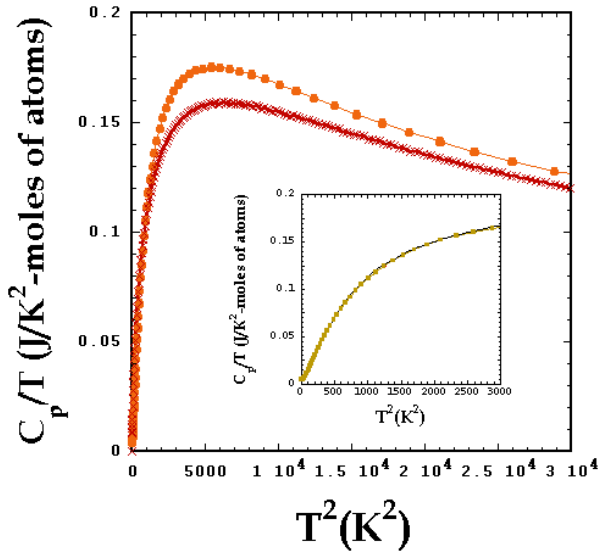}\\
	\caption{(color online).  The heat capacity, ${C}_{p}$, of Yb$_{0.2}$Co$_{4}$Sb$_{12}$ divided by the temperature, \emph{T}, plotted as a function of \emph{T$^{2}$} (\emph{i}) measured \emph{in-situ} (orange $\bullet$) and (\emph{ii}) derived utilizing Eq. 2 from the neutron scattering DOS (red +).  \textit{Inset:} the low temperature section of the heat capacity of the sample and the relevant temperature range for a fit with one Debye and two Einstein modes (from the direct \emph{in-situ} measurement).  The black line is the fit through the data points.}
\end{figure}

\begin{figure}[b]
	\includegraphics[width=0.48\textwidth]{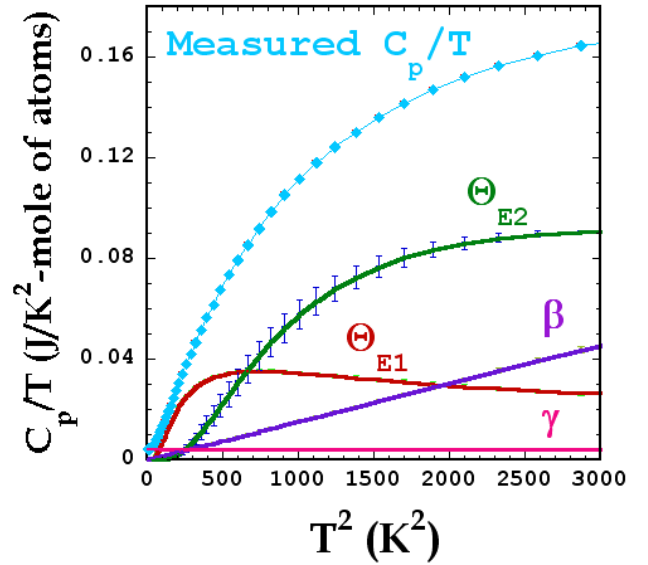}
	\caption{(color online).  The relative contributions of the electronic specific heat \emph{$\gamma$} (in pink), the Debye term, \emph{$\beta$} (in purple), and the two Einstein modes, \emph{${\Theta}_{E1}$} (in red) and \emph{${\Theta}_{E2}$} (in green), obtained from the fit through \emph{C$_{p}$/T} \textit{versus} \emph{T$^{2}$} of the direct \emph{in-situ}-obtained data with Eq. 4.  The total \emph{C$_{p}$/T} is shown in cyan diamonds.  All the data are plotted with their respective errors.  If not visible, the error is smaller than the diameter of the data points.}
\end{figure}

A number of different approaches have been employed in the analysis of heat capacity data of filled skutterudites in the quest for a "rattler" eingenmode.\cite{Mandrus,Keppens,Bauer,Hermann}  In a number of them, the ${C}_{p}$ of the parent compound is subtracted from the one of the filled skutterudite and the resultant curve is fitted with the expected contributions of one or two Einstein modes.\cite{Hermann,Keppens}  These approaches to analyzing heat capacity data in filled skutterudites are based on the assumption that the only difference between parent and filled skutterudites is the one attributed to a rattling mode, but fail to examine other possibilities, such as hybridizations between the vibrational motions of the guest and antimony atoms forming the cage.\cite{Feldman,Veithen}.  We followed the analytic approach used by Mandrus \textit{et al.}\cite{Mandrus} and fitted the \emph{C$_{p}$/T} of the experimental sample Yb$_{0.2}$Co$_{4}$Sb$_{12}$ in the range 6 K $\leq T \leq$ 53.5 K directly with a function of the form:\cite{Mandrus}
\begin{flushleft}
\begin{align}
\begin{split}
f(x) = \gamma + \beta \cdot x 
+ A \cdot {{\Theta}_{E1}}^{2} \cdot {x}^{-\left( 3/2 \right)} \cdot  \frac{{{e^{{\Theta}_{E1}/\sqrt{x}}}}}{{\left( {{e^{{\Theta}_{E1}/\sqrt{x}}}} - 1\right)}^{2}} \\
 + B \cdot {{\Theta}_{E2}}^{2} \cdot {x}^{-\left( 3/2 \right)} \cdot \frac{{{e^{{\Theta}_{E2}/\sqrt{x}}}}}{{\left( {{e^{{\Theta}_{E2}/\sqrt{x}}}} - 1\right)}^{2}}
%\end{flushleft}
\end{split}
\end{align}
\end{flushleft}
, where \emph{x} represents \emph{T$^{2}$}, $\gamma$ is the electronic contribution to the specific heat, $\beta = \frac{12 {\pi}^{4} N {k}_{B}}{5} \cdot {\left( {\Theta}_{D}\right) }^{-3}$ with \emph{N} $-$ the number of host atoms comprising the skutterudite system, and ${\Theta}_{D}$ $-$ the Debye temperature.  ${\Theta}_{E1}$ and ${\Theta}_{E2}$ stand for the Einstein temperatures of the two guest modes necessary for the fit with spectral weights \emph{A} and \emph{B}, respectively. The results of the fit are examplified by the black line through the data points in the inset of Figure 2.

The quality of the fit is significantly better with the inclusion of the second Einstein oscillator: \emph{${\chi}^{2}$} = 1.2 $\times$ 10$^{-3}$ with only \emph{${\Theta}_{E1}$}, and \emph{${\chi}^{2}$} = 2.7 $\times$ 10$^{-5}$ with both \emph{${\Theta}_{E1}$} and \emph{${\Theta}_{E2}$} (see Table I).  The contributions of all those parameters to \emph{C$_{p}$/T} are graphically displayed in Figure 3.

The same series of fits with the above-mentioned models have been performed on the heat capacity derived form the experimental DOS (red '+' plot in Fig. 2).

\begin{table*}
	\begin{tabular}{|c|c|c|c|}
		\hline
		\textbf{Type of Fit} & \textbf{One Debye Mode Only} & \textbf{One-Debye-One-Einstein} & \textbf{One-Debye-Two-Einstein} \\
		\hline
		\emph{$\gamma$} (J/mol-K$^{2}$) & 0.0133 $\pm$ 0.0015 & 0.0054 $\pm$ 0.0003 & 0.0042 $\pm$ 5.2$\times$10$^{-5}$ \\
		\hline
		\emph{$\beta$} (J/mol-K$^{4}$) & 6.85$\times$10$^{-5}$ $\pm$ 1.7$\times$10$^{-6}$ & 3.15$\times$10$^{-5}$ $\pm$ 7.7$\times$10$^{-7}$ & 1.52$\times$10$^{-5}$ $\pm$ 3.6$\times$10$^{-7}$\\
		\hline
		\emph{A} (J/mol-K) & - & 5.356 $\pm$ 0.134 & 1.622 $\pm$ 0.041\\
		\hline
		\emph{${\Theta}_{E1}$} (K) & - & 103.66 $\pm$ 0.89 & 70.04 $\pm$ 0.47\\  
		\hline
		\emph{B} (J/mol-K) & - & - & 8.914 $\pm$ 0.104 \\
		\hline
		\emph{${\Theta}_{E2}$} (K) & - & - & 149.02 $\pm$ 0.90 \\
		\hline
		\emph{${\chi}^{2}$} & 0.049286 & 0.0011986 & 2.6706$\times$10$^{-5}$ \\
		\hline
		\emph{R} & 0.94963 & 0.99881 & 0.99997 \\
		\hline
	\end{tabular}
	\caption{Shown are the results of three different fits on the heat capacity measurement of Yb$_{0.2}$C0$_{4}$Sb$_{12}$ from the \emph{in-situ} data (orange $\bullet$ and line): (1) with only one Debye mode, (2) with one Debye and one Einstein modes, and (3) with a Debye and two Einstein modes, as shown in Fig. 4.  \emph{R} stands for the correlation coefficient of the fit.  The Debye temperature, ${\Theta}_{D}$ was derived from $\beta$ (${\Theta}_{D}$ = 307.65 $\pm$ 2.39 K), and it is comparable to the one obtained in La$_{0.9}$Fe$_{3}$CoSb$_{12}$.\cite{Mandrus}}
\end{table*}

%%%%%%%%%%%%%%%%%%%%%%%%%%%%%%%%%%%%%%%%%%%%%%%%%%%%%%%%%%%%%%%%%%%%%%%%%%%%%%%%%%%%%%%%%%%%%%%%%%%%%%%%%%

\section{Results and Discussion}

The \emph{in-situ} studies of the specific heat in Yb$_{0.2}$Co$_{4}$Sb$_{12}$ (see Fig. 2 orange $\bullet$ data) focused on the low-temperature domain (6 K  $\leq$ \emph{T} $\leq$ 53.5 K) in order to avoid the obscuring influences of anharmonic effects and interference from high-energy optic modes.\cite{Ashcroft}  The data was fit in the specified regime with (i) just one Debye mode, (ii) one Debye and one Einstein modes and (iii) one Debye and two Einstein modes, in order to show the goodnesses-of-fit for all these cases (see Table I).  

Similarly, the heat capacity data obtained from the conversion of the experimental DOS (Eq. 2) was also fitted with Eq. 5 (one Debye and two Einstein modes) in the 6 K $\leq$ \emph{T} $\leq$ 54 K range, in order to check the consistency of the data in Table I.  The following results were obtained: $\gamma'$ = 0.0169 $\pm$ 0.0004 J/mol-K$^{2}$; $\beta$ = 1.44 $\times$ 10$^{-5}$ $\pm$ 5.7 $\times$ 10$^{-7}$ J/mol-K$^{4}$; \emph{A} = 1.440 $\pm$ 0.055 J/mol-K; ${\Theta}_{E1}$ = 60.616 $\pm$ 1.020 K; \emph{B} = 6.309 $\pm$ 0.156 J/mol-K; ${\Theta}_{E2}$ = 146.53 $\pm$ 2.020 K; ${\chi}^{2}$ = 7.3886 $\times$ 10$^{-6}$ and \emph{R} = 0.99995.  By comparison, the fit of the same data with just one Einstein mode resulted in ${\chi}^{2}$ = 0.00035394 and \emph{R} = 0.99776.

The \emph{in-situ} heat capacity measurement is susceptible to error introduced by parasitic heat losses, blackbody radiation effects, while the DOS conversion is sensitive to background subtraction and the resolution of the CNCS.  However, the remarkable agreement between the fitting coefficients obtained from the fits with two Einstein modes on both the \emph{in-situ} and DOS-derived heat capacity data, compared to those of any other model used, gives a reliable consistency check on the quality of the data and the theory presented.

It is evident, as shown in Fig. 2 and Table I, that especially at \emph{T} $>$ 30 K the data could not be accounted for by fewer than two incoherent oscillator modes superimposed on the electronic and coherent phonon contributions to the specific heat.  The excellent correlation coefficient for the One-Debye-Two-Einstein mode scenario fit of the \emph{in-situ} measurement (\emph{R} = 0.99997) produced a very reliable fit, suggesting that in order to properly take account of that data we need a minimum of two "rattler" modes (see Table I).

\begin{figure}
 	\includegraphics[width=0.48\textwidth]{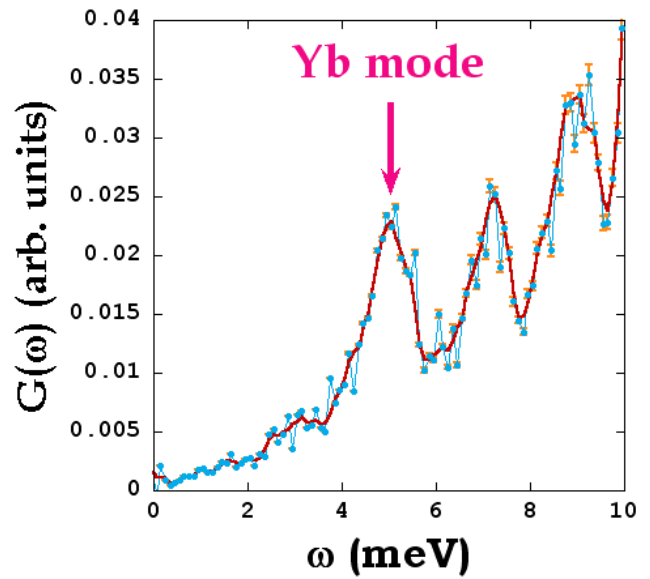}
	\caption{(color online).  Close-up of the experimentally-measured density of states of Yb$_{0.2}$Co$_{4}$Sb$_{12}$ (the red line through the data points is a guide for the eye) exhibits a clearly-defined mode beyond the experimental resolution at $\omega \approx$ 5 meV, which is attributed to the dynamics of Yb, as shown in Fig. 1(b).}
\end{figure}

\begin{figure}
 	\includegraphics[width=0.46\textwidth]{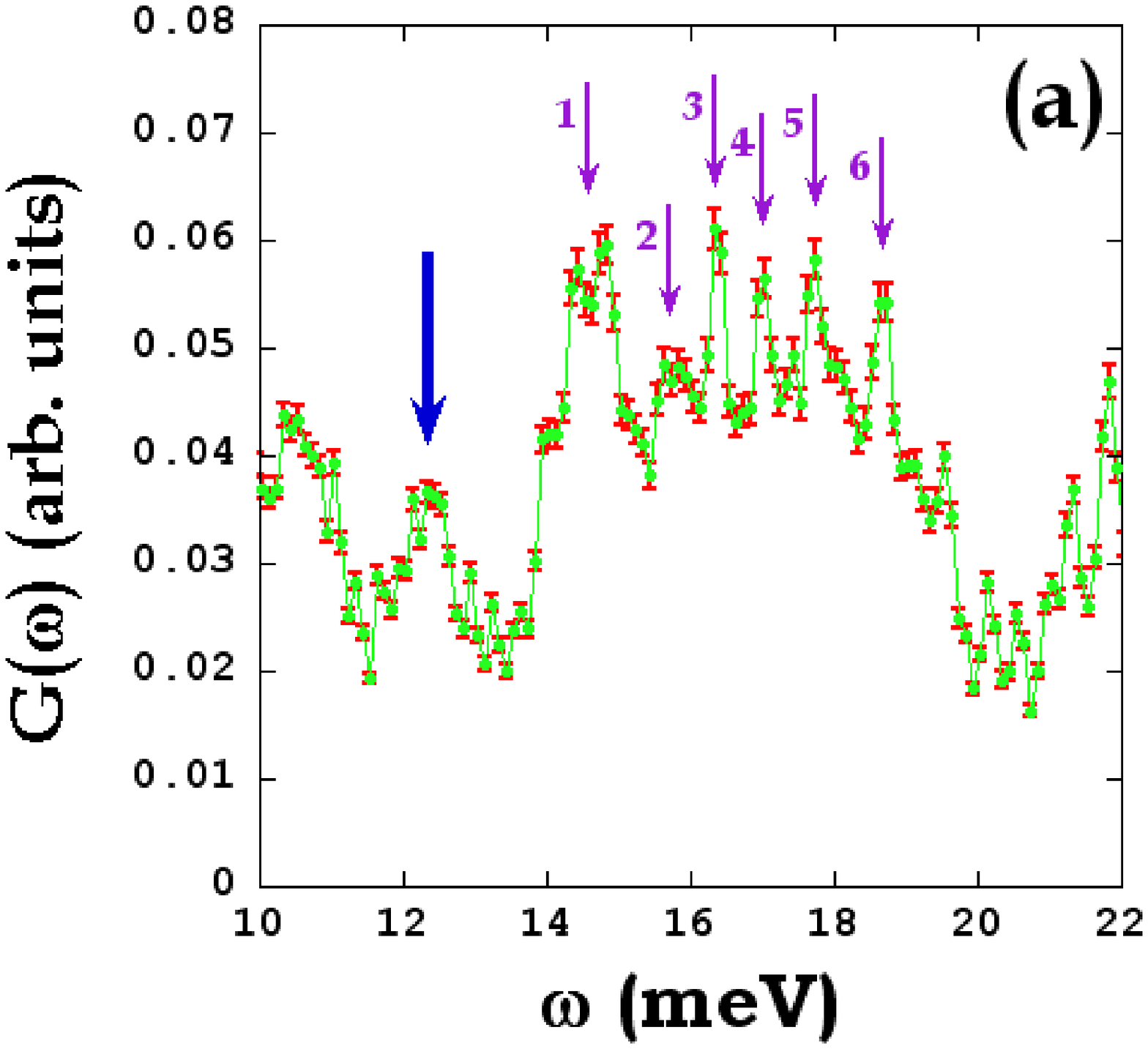}\vspace{0.5cm}
 	\vspace{0.5cm}
  	\includegraphics[width=0.46\textwidth]{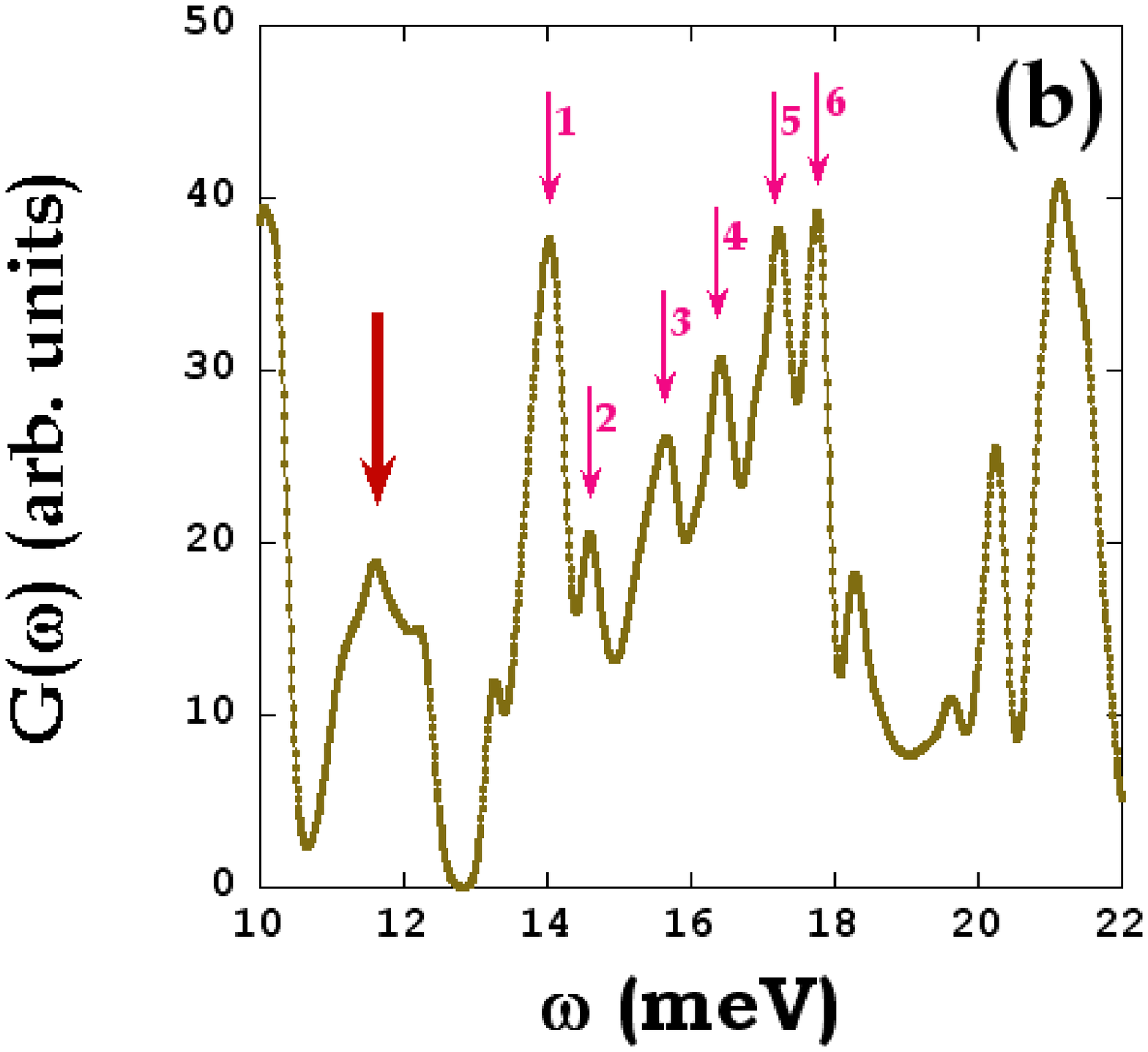}
	\caption{(color online).  Shown are close-ups of (a) the experimentally-measured $G(\omega)$ of Yb$_{0.2}$Co$_{4}$Sb$_{12}$ and (b) DFPT calculation of $G(\omega)$ of Yb$_{0.25}$Co$_{4}$Sb$_{12}$, both at higher transfer energies (13 - 19 meV), where a number of sharp features are resolved in the experimental data.  The large blue and red arrows identify the locations of the peak ($\omega \, \approx$ 12.4 meV) in the DOS of Yb$_{0.2}$Co$_{4}$Sb$_{12}$, which is identified with the second Einstein mode, ${\Theta}_{E2}$.  The small purple and pink arrows identify the locations of local modes at intermediate transfer energies ($\omega \, \approx$ 10 -- 22 meV) in the experimental data and DFPT calculation, respectively.}
\end{figure}

The density of states \emph{G($\omega$)} obtained from the inelastic neutron time-of-flight measurement on Yb$_{0.2}$Co$_{4}$Sb$_{12}$ reveals the presence of low-lying phonons ($\omega \, \approx5 - 25$ meV), separated from high-energy ones ($\omega \, \approx 27 - 35$ meV) (see Fig. 1).  As shown in Fig. 1, there is an excellent agreement between features observed in our experimental data (Fig. 1(a)) with those predicted by the density functional perturbation theoretical (DFPT) calculation on Yb$_{0.25}$Co$_{4}$Sb$_{12}$ (Fig. 1(b)).\cite{John}  The modes observed between 27 meV and 35 meV are shown to be mainly due to vibrations of the Co atoms and are in the same energy range as "iron modes" observed in the Fe-based skutterudite (Ce,La)Fe$_{4}$Sb$_{12}$.\cite{Koza}

A significant local maximum from the calculations of the phonon DOS is a low energy peak at $\omega \approx$ 5.52 meV, which is attributed to vibrations of the Yb atoms (see Fig. 1(b)).  A corresponding feature to the latter also appears in the measured DOS, as shown in Figs. 1(a) and 4, namely, a peak at 5.0 meV, which is resolved and well-separated from the band of Sb vibrations between 7 meV and 23 meV (the high resolution of the CNCS instrument is readily visible in $G(\omega)$ \emph{vs.} $\omega$ close-ups in two separate $\omega$ ranges, as shown in Figs. 4 and 5(a)).  The results from the fit with two Einstein oscillators of the \emph{in-situ} specific heat measurement (see Figs. 2 and 3) corroborate the presence of a low-energy Einstein mode, \emph{${\Theta}_{E1}$} = 70 K (${\omega}_{E1}$ = 6.0 meV), which corresponds to the measured and calculated values for the Yb peak at $\omega \approx$ 5 meV in $G(\omega)$ \emph{vs.} $\omega$ (see Figs. 1 and 4).  

In previous heat-capacity studies on filled skutterudites a good consistency was observed between the weight of the first Einstein oscillator mode (\emph{A}) with the theoretical prediction that each oscillator heavy element atom should contribute 1.47 J/K-mole of oscillator atoms.\cite{Mandrus,Keppens}  In the latter papers the authors cite \emph{A} = 1.21 J/K-mole in La$_{0.9}$Fe$_{3}$CoSb$_{12}$.\cite{Mandrus,Keppens}  Ironically, in the case of Yb$_{0.2}$Co$_{4}$Sb$_{12}$, we would expect \emph{A} to be 0.308 J/K-mole, whereas the fitted value is shown to be 1.62 J/K-mol -- a difference by a factor of 5.25.  The apparent similarity of \emph{A} in Yb$_{0.2}$Co$_{4}$Sb$_{12}$ and La$_{0.9}$Fe$_{3}$CoSb$_{12}$, despite the fact that the former has 4.5 times fewer “rattler” atoms than the latter, suggests that even the peak at 5 meV (\emph{${\Theta}_{E1}$} mode) could be a consequence of a local change in the Co$_{4}$Sb$_{12}$ lattice.  If one assumes that the nearest neighbors of Yb become loosely bound to the skutterudite lattice due to the filler, then their added degrees of freedom should account for the much greater \emph{A} observed in Yb$_{0.2}$Co$_{4}$Sb$_{12}$.     

The pronounced appearance of the eigenmode at 5 meV in Yb$_{0.2}$Co$_{4}$Sb$_{12}$ is attributed to the large neutron flux and improved resolution of the CNCS, but also to a large extent $-$ to the large neutron scattering cross section of ytterbium ($\sigma$(Yb) = 23.4 bn, $\sigma/M$ = 0.13).  The Yb mode is characterized by an energy range similar to other filler modes observed in a number of Ce, La and Tl-filled skutterudite systems\cite{Hermann,Koza}, and exhibits a striking resemblance to the results of a recent high resolution inelastic neutron scattering experiment and theoretical DOS calculation on another Yb-filled system, YbFe$_{4}$Sb$_{12}$.\cite{Koza2}  The latter system exhibits two separate filler modes, which manifest as distinct peaks at 4.9 meV and 5.7 meV in the experimental data, which are also seen in the theoretical calculations of Koza \textit{et al.}\cite{Koza2}.  These are to be contrasted with our own results, which attest to a single peak in Yb$_{0.2}$Co$_{4}$Sb$_{12}$, both theoretically and experimentally within resolution limits.  We attribute this distinction between the guest modes of the two samples to differences in the curvatures of the Yb binding potentials in the FeSb$_{3}$ and CoSb$_{3}$ cages, respectively.  

Furthermore, the specific heat measurement of Yb$_{0.2}$Co$_{4}$Sb$_{12}$ also seems to suggest the presence of another local mode at higher energy transfers ($\omega \, >$ 5 meV), beyond the Yb peak.  In addition to \emph{${\Theta}_{E1}$}, a second Einstein mode, \emph{${\Theta}_{E2}$} = 149.02 K, (\emph{$\omega$} = 12.8 meV) is also accounted for from the functional fit through the specific heat data (see Fig. 2), and coincides with a peak in the DOS of Yb$_{0.2}$Co$_{4}$Sb$_{12}$ at 12.4 meV (see Fig. 5(a) and 5(b)).  The presence of \emph{${\Theta}_{E2}$} seems consistent with the two Einstein modes found in a number of other systems and has been related to the phonon glass.\cite{Mandrus,Keppens,Bauer}  Upon closer inspection of the contributions of \emph{${\Theta}_{E1}$} and \emph{${\Theta}_{E2}$} to the total \emph{${C}_{p}/T$}, \emph{${\Theta}_{E2}$}'s part exceeds \emph{${\Theta}_{E1}$}'s throughout the high temperature range, whereas below $T \approx$ 27 K, \emph{${\Theta}_{E1}$} begins to dominate (see Fig. 3).\cite{Mandrus}  However, the effective contributions of the two Einstein modes to the lattice thermal conductivity would be modified by the power-law frequency dependence of the attenuation of sound waves
 $\alpha$ (${\alpha}^{shear \; waves} \propto {\omega}^{2}$ and ${\alpha}^{longitudinal \; waves} \propto {\omega}^{4}$), as well as the absolute temperature, \emph{T}, and the elastic constants of the system,\cite{Klemens} which could be obtained from a velocity of sound measurement on filled skutterudite systems.  
 
Another remarkable finding in our experiment is the presence of sharp modes (\emph{van Hove singularities}) in the 14 meV $\leq$ $\omega$ $\leq$ 20 meV range (see Figs. 1(a) and 5(a)), which the DFPT calculation shows (Fig. 5(b)).  These are mainly attributed to vibrations of Sb atoms (see Fig. 1(b)).  The 14 -- 20 meV energy transfer range is distinguished by 6 sharp peaks (shown in Fig. 5), and the one listed as "\emph{1}" occurs at an energy transfer where Sb and Co peaks are coincident in the DOS of Yb$_{0.2}$Co$_{4}$Sb$_{12}$ (see Fig. 1(b) and Fig. 5).  This suggests that peak "\emph{1}" may be due to a hybridization of atomic degrees of freedom, as suggested by Veithen and Ghosez and Feldman \emph{et al.}, whose calculations of the dispersion curves of CoSb$_{3}$ exhibit a large number of nearly dispersionless optic modes that result in sharp features in the calculated phonon DOS.\cite{Veithen,Feldman}  What is surprising in our study is that we observed these sharp features, whereas other such experiments on CoSb$_{3}$ show only a broad peak in this energy range.\cite{Hermann}  

\begin{figure}[t]
	\includegraphics[width=0.46\textwidth]{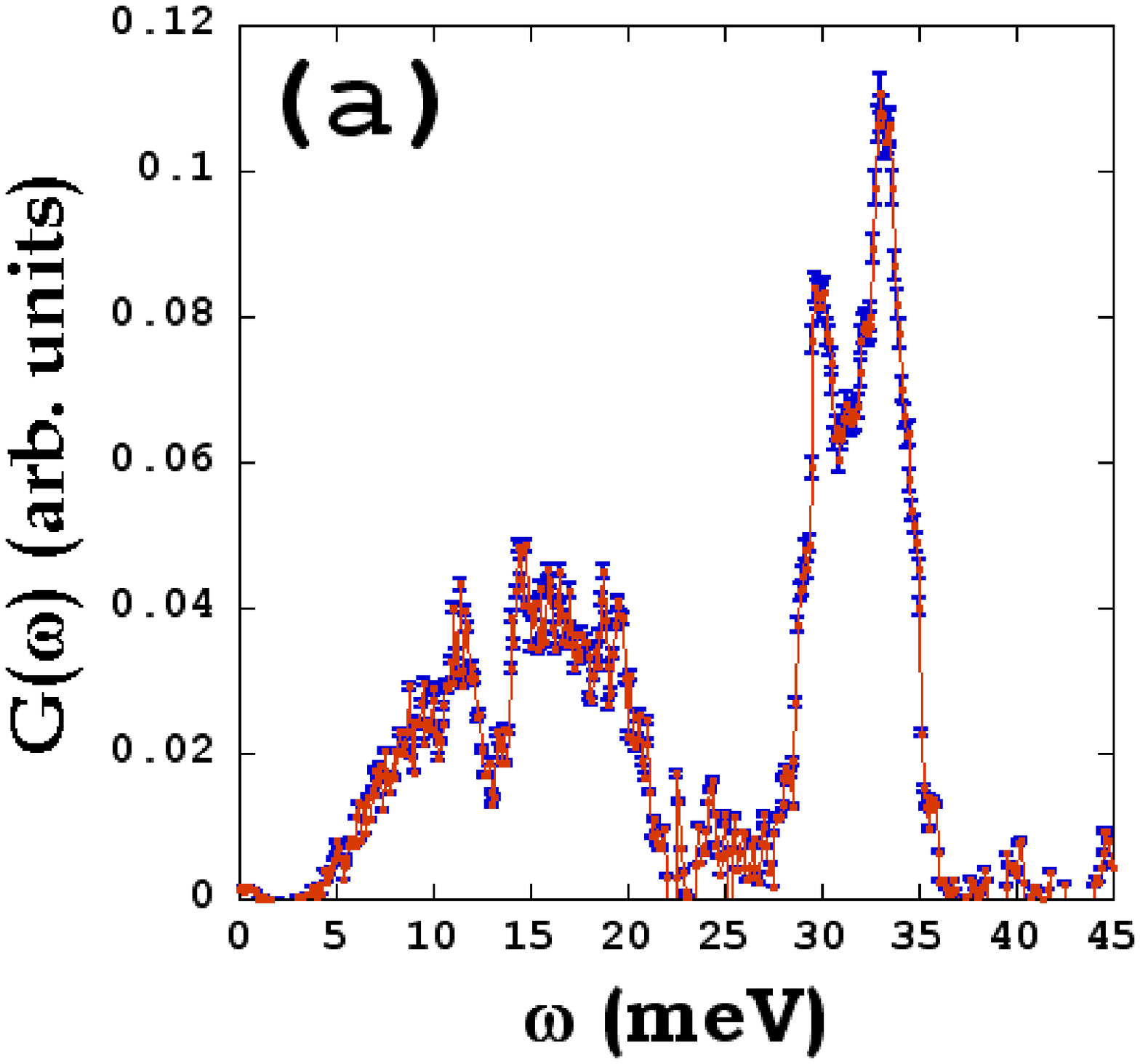}\\
	\vspace{1cm}
	\includegraphics[width=0.46\textwidth]{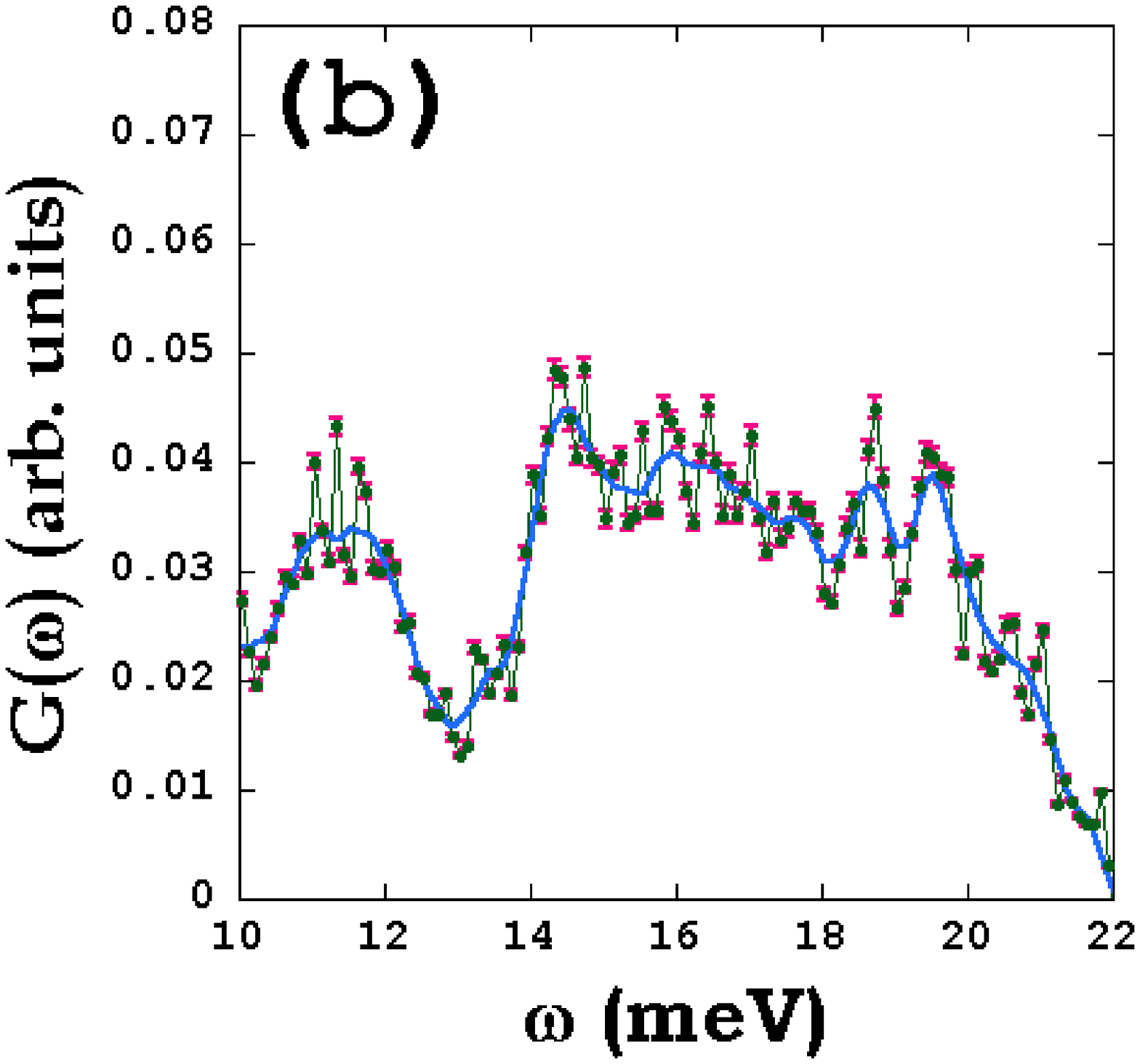}
	\caption{(color online).  (a) The density of states of Ce$_{1.05}$Fe$_{4}$Sb$_{12.04}$ shows the same qualitative behavior as the results of a previous experimental and theoretical investigation of CeFe$_{4}$Sb$_{12}$.\cite{Koza}  (b) The DOS of Ce$_{1.05}$Fe$_{4}$Sb$_{12.04}$ zoomed in on the same energy and amplitude scales as the corresponding DOS for Yb$_{0.2}$CO$_{4}$Sb$_{12}$ failed to show well-defined dispersionless peaks.  The blue line is a guide for the eye.}
\end{figure}

To further study the proposed guest-host coupled dynamics of filled skutterudites, we also measured the neutron-weighted density of states of a related skutterudite prepared in the same way: Ce$_{1.05}$Fe$_{4}$Sb$_{12.04}$.  $G(\omega)$ of Ce$_{1.05}$Fe$_{4}$Sb$_{12.04}$ is shown in Fig. 6(a) and compares favorably with the previously-measured DOS of CeFe$_{4}$Sb$_{12}$.\cite{Koza}  A broad feature is present in the energy range 14 meV $\leq$ $\omega$ $\leq$ 20 meV, and appears quite distinct from the sharp peaks observed in the Yb$_{0.2}$Co$_{4}$Sb$_{12}$ DOS (see Figs. 5(b) and 6(b)).  Even though Ce$_{1.05}$Fe$_{4}$Sb$_{12}$ seems to have a local maximum in the DOS centered around 14.3 meV (very faint bump), it appears to be less significant than the corresponding one (a peak at 14.7 meV marked with a blue arrow in Fig. 5(a)) in the experimentally-measured DOS of Yb$_{0.2}$Co$_{4}$Sb$_{12}$.  ${\omega}_{peak}$ is approximately 14 meV by the calculation and is marked with a red arrow in Fig. 5(b) (see Figs. 5(a), 5(b) and 6(b)).  On the other hand, the presence of ytterbium cannot be exclusively credited with the appearance of dispersionless modes in the intermediate energy range in the DOS of Yb$_{0.2}$Co$_{4}$Sb$_{12}$, since the density of states of YbFe$_{4}$Sb$_{12}$, a different Yb-filled antimonide, shows a relatively broad feature in the same energy transfer range.\cite{Koza2}  

For the most part, Fig. 6(b) shows noisy data which indicate that either Ce$_{1.05}$Fe$_{4}$Sb$_{12.04}$ does not exhibit dispersionless modes or that even if they were to exist they could not be resolved by the experiment.  The broad noisy shoulder in CeFe$_{4}$Sb$_{12.04}$ has also been observed previously by Koza \emph{et al.}, with no additional features standing out.\cite{Koza} 

The DOS comprises the sum of the modes of the system over all reciprocal lattice vectors at a given energy exchange.  Therefore, our experimental DOS represents a superposition of many different bands and the relative contributions of each of them.  Thus, the important questions are not only \emph{what is the scattering cross section of Yb versus Ce}, since in the intermediate energy-transfer regime the theoretical calculations do not expect ytterbium or cerium to contribute to the density-of-states, but rather -- \emph{do dispersionless modes exist in the Ce-filled compound} and \emph{how do they compare with the contributions of coherent modes to the density-of-states}.  We attribute the noisy broad shoulder with no distinct peaks in Ce$_{1.05}$Fe$_{4}$Sb$_{12.04}$ to a smaller relative amplitude of incoherent modes compared to Yb$_{0.2}$Co$_{4}$Sb$_{12}$.  

The effects of the specific filler on the specific cage must be therefore considered.  Yb$_{0.2}$Co$_{4}$Sb$_{12}$ seems to be a very special system, since it defies the traditional "rattling mode scenario" and seems to suggest a more complicated system than the ones previously described.

The discrepancy between the theoretical calculation of spectral weight \emph{A} and the experimentally measured one and the differences among densities-of-states between Yb$_{0.2}$Co$_{4}$Sb$_{12}$ and other similar systems suggest that a possible cause for the emergence of local modes in the 10 -- 20 meV energy transfer range are incoherent lattice vibrations of Yb nearest neighbors in the Yb$_{0.2}$Co$_{4}$Sb$_{12}$ cage, since at those energies Yb is not expected to contribute directly.  The DFPT calculation fully relaxes the filled structure, calculates the dynamic matrix at first and then performs the phonon spectrum calculation. Thus, the DFPT calculations do include the effect of structure change, such as the structure relaxation and bonding strength change, due to the filling. In other words, these effects do contain the "coupling" of Yb rattling and local Sb vibrational modes in an implicit way.
But no phonon-phonon coupling is explicitly included.  In principle, there could be a change in the phonon DOS in the 14 -- 20 meV regime as a result of Yb filling because the Sb configuration of the cages does change in comparison with the unfilled systems.  

Altered lattice properties due to the presence of a heavy earth element could lead to localized "pocket" modes which manifest themselves as dispersionless modes at $\omega$ = ${\omega}_{E2}$ = 12.4 meV, as well as in the 14 -- 20 meV regime,\cite{Sandusky,Grant} i.e., low-energy modes appearing in the gap of a phonon-gapped system resulting from a pseudo-vacancy at an impurity site.  Such modes result from the displacements of lighter atoms, such as Sb, which are nearest-neighbor atoms to the Yb filler.  The vacancy which is left as a result of the displacement would be a mode associated with atoms moving in and out of these \emph{quasi}-holes in the lattice.

The appearances of a second Einstein mode (${\Theta}_{E2}$) and other sharp modes in the DOS of Yb$_{0.2}$Co$_{4}$Sb$_{12}$ in the 14 meV $\leq$ $\omega$ $\leq$ 20 meV range (where ytterbium is shown to \emph{not} contribute to the phonon density-of-states) are attributed to altered properies of the skutterudite cage due to the addition of ytterbium.  Given the high \emph{ZT} of the filled skutterudites over their unfilled parents, the above-mentioned coincidence of dispersionless modes from the phonon density of states with a second Einstein mode indicates that the scattering of heat-carrying phonons occurs at a variety of energy scales and could also be owed to a novel mechanism which merits further research and investigation.

\section{Conclusions and Suggestions}

Our results have shown a pronounced peak in the phonon DOS of Yb$_{0.2}$Co$_{4}$Sb$_{12}$ at \emph{$\omega$} $\approx$ 5 meV, established from inelastic neutron time-of-flight scattering, specific heat measurements and theoretical calculations.  This is interpreted as a "rattler" mode of Yb atoms vibrating in the cage of the CoSb$_{3}$ host lattice.  

Novel contributions of this work include the observations of the second Einstein oscillator in the DOS of Yb$_{0.2}$Co$_{4}$Sb$_{12}$ at ${\omega}_{E2}$ = 12.4 meV, and the dispersionless modes in the \emph{$\omega$} = 14 -- 20 meV energy-transfer regime.  Both are believed to arise as a consequence of ytterbium filling.  It appears that incoherent modes from the cage itself are induced by the addition of Yb and these also participate in heat dissipation.  

Future acoustic experiments aimed at measuring differences between the elastic constants of filled and parent cobalt antimonides as a function of frequency should be performed in order to understand the significance of each "rattler" mode to the total lattice thermal conductivity, ${\kappa}_{L}$.\cite{Klemens}  In addition, future experiments aimed at measuring the DOS of filled cobalt antimonides as a function of filler concentration should try to assess the role of the corresponding filler on the number of intermediate energy modes, their amplitudes and possible energy shifts in the latter.  

The further understanding of intermediate energy modes will most likely illuminate the importance of filler atoms in filled skutterudite thermoelectrics and allow the engineering of materials with higher dimensionless figures of merit.

\acknowledgements
The work at Brookhaven National Laboratory was supported by the Office of Science, U.S. Department of Energy, under Contract No. DE-AC02-98CH10886.  M.E.M.'s work was performed under the auspices of the U.S. Department of Energy by Lawrence Livermore National Laboratory under Contract No. DE-AC52-07NA27344.  This work was partly supported by National Basic Research Program of China (Contract No. 2007CB607500), National Natural Science Foundation of China (Contract No. 50821004), National Science Foundation for Distinguished Young Scholars of China (Contract No. 1150825205).  This Research at Oak Ridge National Laboratory's Spallation Neutron Source was sponsored by the Scientific User Facilities Division, Office of Basic Energy Sciences, U. S. Department of Energy.  I.K.D. wishes to thank Vyacheslav Solovyov for stimulating discussions and critical reading of the manuscript.


\begin{references}
\bibitem{John}
The reason for the slight differences between the compositions of ytterbium in Yb$_{\delta}$Co$_{4}$Sb$_{12}$ ($\delta$ = 0.2 and 0.25 for the experimental sample and the theory, respectively) chosen for the experiment and the theory are due to the fact that the experimental work was carried prior to the PALD calculation, for the purposes of which a 2$\times$2$\times$2 supercell was constructed for the Co$_{4}$Sb$_{12}$, i.e., Co$_{32}$Sb$_{96}$ in the actual simulation.  Subsequently, 2 out of the 8 voids in the supercell are filled with Yb atoms and for computational convenience Yb$_{0.25}$ has been chosen instead of Yb$_{0.2}$.  However, as seen in the comparison, this slight compositional variation does not produce visible qualitative differences.
\end{references}
\end{document}